\documentclass[12pt]{article}
\usepackage{amsmath}

\begin{document}

\title{SU(2) chiral Yang--Mills model on a lattice. Witten anomaly eaten by
the lattice}
\author{\textbf{Corneliu Sochichiu} \\
Bogoliubov Laboratory \\
of Theoretical Physics\\
JINR\\
141980 Dubna\\
Moscow Region\\
Russia\\
e-mail: sochichi@thsun1.jinr.dubna.su}
\date{}
\maketitle

\begin{abstract}
Using a regularisation of the chiral SU(2) Yang--Mills model by an infinite 
number of Pauli--Villars fields the doubler free gauge invariant lattice
model is constructed. Its continuum limit provides the recently proposed
mechanism for cancelation of Witten anomaly.
\end{abstract}

Formulation a model containing chiral fermions on a lattice is a long
standing problem due to the phenomenon of fermion doubling \cite{w}. It was
proposed by Wilson to cure the doubling by addition of an extra term to the
naive lattice action which produces large masses of the order of cut off to
doubling states. This eliminate doublers in continuum limit but it destroys
the chiral invariance as well. It also is the source of chiral anomaly in
the continuum limit. In some cases, however, such as QCD in the limit of
zero mass $u$ and $d$ quarks, SM or other nonanomalous models this
destruction is too ``strong''. In this case one still needs 
additional ``fine tuning'' of the model in order to approach the 
chiral limit. The counter-terms needed for the ``fine tuning'' are 
not known {\it a priori}, and require a careful (non-perturbative) 
computation (for a recent review see e.g.  \cite{lw}). From the other 
hand, no local lattice formulation preserving the non-anomalous 
chiral symmetry for finite lattice spacings is known in present.

Moreover, there exists a theorem due to Nielsen and Ninomiya \cite{nn} which
states that under certain circumstances such as locality, hypercubic and
gauge invariance and hermicity there are no chiral fermions on a lattice.

From the other hand among the nonanomalous models there exist ones given by
the real representations of gauge groups. For such models one can formulate
a lattice description satisfying all conditions of Nielsen--Ninomiya theorem
but with no doubling in continuum limit. This is possible because both
chiralities in real representations are equivalent. For example for a model
describing a chiral fermion in $n$-dimensional representation of $SO(n)$ one
can write down the gauge invariant Wilson term as follows
\[
\frac 12a\psi ^{\text{T}}C\Delta \psi +\text{h.c.},
\]
where $a$ is the lattice spacing, $C$ is the Dirac charge conjugation
matrix, $\Delta $ is the lattice Laplace operator and ``T'' stands for
transposed spinor.

In case of $SU(2)$ model described by the Lagrangian
\begin{eqnarray}
\mathcal{L} &=&-\frac 1{4g^2}F_{\mu \nu }+\overline{\psi }_{+}\widehat{
\nabla }\psi _{+}  \label{1} \\
P_{-}\psi _{+} &=&0,\ P_{\pm }=\frac 12(1\pm \gamma _5)  \nonumber
\end{eqnarray}
where $\psi _{+}$ belongs to the fundamental representation of $SU(2)$ one
can construct a field
\begin{equation}
\overline{\psi }_{+}^c=\psi _{+}^{\text{T}}C\cdot i\sigma _2,  \label{2}
\end{equation}
where $C$ is the same Dirac charge conjugation matrix and $\sigma _2$ is
Pauli $\sigma _2$-matrix (here it plays the role of $SU(2)$ charge
conjugation matrix). One can see that $\overline{\psi }_{+}^c$ defined in
such a manner has the same chirality but transforms by conjugate action of
the gauge group as initial field $\psi _{+}$. One can try to write down the
gauge invariant Wilsonian term in analogy with $SO(n)$ case but since the
product $C\cdot i\sigma _2$ is symmetric this term will be identically zero
due to anticommuting properties of the fermionic field. As we will see later
it is connected with Witten anomaly. If the number of fermionic fields were
even there would be no problem since one can take the crossing product of
different fields solving the problem. But one can re-define the fields in
this case to show that in fact one deals with half number of Dirac fermions.

In what follows we will adopt a different strategy. In continuum one can
regularize this model by introducing an infinite number of Pauli--Villars
(PV) fields \cite{fs1} as follows
\begin{equation}
\mathcal{L}_{\text{reg}}=\overline{\psi }_{+}\widehat{\nabla }\psi
_{+}+\sum_{r=1}^\infty \overline{\psi }_r(\widehat{\nabla }+Mr)\psi _r
\label{4}
\end{equation}
where $\psi _r$ are Dirac spinors with Grassmannian parities $\left(
-\right) ^r$. Regularizations of this type were used to prove the gauge
invariance of the continuum limit of lattice models with Wilson fermions
\cite{fs2}, Smit--Swift model \cite{sl1} and for chiral gauge invariant
lattice formulation of the $SO(10)$ unified model with SLAC fermions \cite
{sl2}\footnote{An alternative way may consist in considering second 
scale regularization using a coarser lattice, for a recent review see 
\cite{pilar}, and references therein.}.  The same regularization was used 
recently for the construction of the representation for global Witten 
anomaly in continuous $SU(2)$ chiral YM model \cite{sl3}.

Let us return to the lattice model. We will introduce the following
regularized lattice action
\begin{eqnarray}
\mathcal{L}_{reg} &=&\overline{\psi }_{+}\left( \widehat{D}+\frac 1{2\Lambda
^2}\{\widehat{D},\Delta \}\right) \psi _{+}+  \label{5} \\
&&\ \sum_{r=1}^\infty \left\{ \overline{\psi }_r\left( \widehat{D}+\frac
1{2\Lambda ^2}\{\widehat{D},\Delta \}\right) \psi _r+Mr\overline{\psi }
_r\psi _r\right\}  \nonumber
\end{eqnarray}
where $\widehat{D}$ is the naive lattice Dirac operator given by
\begin{equation}
\left( \widehat{D}\psi \right) (x)=\frac 1a\sum_\mu \gamma _\mu \left( U_\mu
(x)\psi (x+ae_\mu )-\psi \left( x\right) \right)  \label{6}
\end{equation}
here $a$ is the lattice spacing and $e_\mu $ is the unite vector in the $
x_\mu $ direction, $U_\mu (x)$ is the lattice gauge field corresponding to
the link connecting $x$ and $x+ae_\mu ,$ $\Delta $ in the eq.(\ref{5}) is
the lattice Laplace operator the same (up to an extra factor $a^{-1}$) as
used in the Wilson term and $M$ and $\Lambda $ are new cut off parameters.
The action (\ref{5}) generates the following Feynman rules
\begin{eqnarray}
s_r &=&\frac{\sum_\mu \gamma _\mu P_\mu (p)+Mr}{P^2(p)+M^2r^2},\quad r\neq 0
\label{prop} \\
s_0 &=&P_{+}\frac{\sum_\mu \gamma _\mu P_\mu (p)}{P^2(p)},
\end{eqnarray}
for propagator and
\begin{equation}
V_{\mu _1\dots \mu _n}^n\sim \frac{\partial ^n}{\partial k_\mu {}_1\dots
\partial k_\mu {}_n}\sum_\alpha \gamma _\alpha P_\alpha (k)|_{k=\frac
12(p+q)},  \label{vert}
\end{equation}
for the vertices. In the above equations $P_\alpha (p)$ is given by
\begin{equation}
P_\alpha (p)=\frac 1a\sin p_\mu a\left( 1+\frac 1{\Lambda ^2a^2}\sum_\alpha
(1-\cos p_\alpha a)\right) .  \label{p}
\end{equation}
If $n=1$ in the vertex we call it a simple one and if $n>1$ we call it
multiple one.

Let us show that in the framework of the perturbation theory one will have
the contribution of doubling states vanishing in the continuum limit if the
cutoffs $M$ and $\Lambda $ go to infinity as $\frac 1{\sqrt{a}}$.

To analyze the continuum limit let us show as an example calculation of a
diagram with $n$ simple external legs. Generalization to multiple legs is
straightforward. The Feynman integral corresponding to the diagram looks as
follows
\begin{eqnarray}
I_{\mu _1...\mu _n}^{i_1...i_n}\left( k_1,...,k_n\right) &=&\tau
^{i_1...i_n}\int_{-\frac \pi a}^{\frac \pi a}\frac{d^4p}{(2\pi )^4}\text{tr}
P_{+}V_{\mu _1}(p,p+k_1)s(p+k_1)...  \label{7} \\
&&...V_{\mu _n}(p-k_n,p)s(p)+  \nonumber \\
&&\sum_{r=1}^\infty (-)^r\text{tr}V_{\mu 
_1}(p,p+k_1)s_r(p+k_1)....V_{\mu _n}(p-k_n,p)s_r(p)  \nonumber 
\end{eqnarray}
where $\tau ^{i_1...i_n}$ is the group factor.

Function $P_\alpha (p)$ given by the eq.(\ref{p}) near the edge of the
Brillouin zone behaves as follows
\begin{equation}
P_\alpha (p')\approx \frac{2\xi }{\Lambda ^2a^2}p'_\alpha +O(1),  \label{edge}
\end{equation}
where $2\xi =\sum_\mu (1+(-)^{n_\mu })$, and quantities $n_\mu$ and 
$p'$ are defined by the following
\begin{equation}
-\frac{\pi}{2a}<p'_\mu \equiv p_\mu-\frac{\pi}{a}n_\mu \leq 
\frac{\pi}{2a} \nonumber \end{equation}

At the center of the Brillouin zone $ P_\alpha 
(p)$ has the usual expansion \[ P_\alpha (p)\approx p_\alpha +...  \] 
where dots stay for terms vanishing as $a^2$ and faster.

The propagator and the vertex at the edge of the Brillouin zone behaves like
this
\begin{eqnarray}
s_r(p') &\approx &\frac 1{\frac{2\xi }{\Lambda 
^2a^2}\hat{p'}-Mr},\qquad \hat{p'}\equiv {p'}_\mu \gamma_\mu
\label{12} \\ V_\mu &\approx &\frac{2\xi 
}{\Lambda ^2a^2}\gamma _\mu .  \end{eqnarray}

One can see that if there were no PV regularization one would have near the
edge of the Brillouin zone almost the same integrand (the factors $\frac{
2\xi }{\Lambda ^2a^2}$ cancel out) except the mass term which here vanishes
as $\frac{\Lambda ^2a^2}{2\xi }M\rightarrow 0.$

Since one have the PV regularization the situation changes dramatically.
Indeed, the leading part of the integrand behaves like (hereafter we 
omit primes for $p'$),
\begin{eqnarray} &\sim &\sum_r(-)^r\frac 
1{(P^2(p)+M^2r^2)^n}\left( P^{\prime }(p)\right) ^nP^n(p)\sim  
\label{a13} \\ &\sim &\left( \frac \partial {\partial P^2(p)}\right) 
^{n-1}\frac \pi {MP^2(p)\sinh (\pi P(p)/M)}\left( P^{\prime 
}(p)\right) ^nP^n(p)  \nonumber \end{eqnarray} The regularization 
produces exponential cut off for zones where $P(p)\gg M\sim \frac 
1{\sqrt{a}}$. As lattice spacing goes to zero and $\Lambda \sim M\sim 
\frac 1{\sqrt{a}}$ to infinity one will have the cut off ``starting'' 
from values of $p\sim \frac 1{\sqrt{a}}$ and ``finishing'' by the values of
$p$ approaching the Brillouin zone edge as $|p-\frac \pi a|\sim \sqrt{a}$.
Indeed from eqs (\ref{edge},\ref{a13}) one can see that for lattice spacings
$a$ small enough the asymptotic behavior of the integrand near the Brillouin
zone edge is as follows
\begin{equation}
\sim \left( \frac{\Lambda ^2a^2}{2\xi }\frac \partial {\partial p^2}\right)
^{n-1}\frac \pi {M\left( \frac{2\xi }{\Lambda ^2a^2}\right) ^2p^2\sinh
\left( 2\xi \pi ^2p/M\Lambda ^2a^2\right) },  \label{a14}
\end{equation}
while in the center of the zone integrand behaves like
\begin{equation}
\sim \left( \frac \partial {\partial p^2}\right) ^{n-1}\frac \pi {Mp^2\sinh
\left( \pi ^2p/M\right) }.  \label{a15}
\end{equation}

From the above equations one can see that the integral over central zone
tends to the correct continuum value (after extraction of UV divergencies)
while integration over the strip close to the edge of the Brillouin zone
decrease with $a$ going to zero as $\sim a^2$. In the domain of momenta
between these two extrema i.e. when $p$ is departed from the center by more
then value of $M$ and from the edge by $M\Lambda ^2a^2\sim \sqrt{a}$ the
integrand decays exponentially and its contribution to the integral vanishes
in the continuum limit.

In fact to be rigorous in the analysis of the behavior of the Feynman
integral near the edge of the Brillouin zone one cannot neglect the external
momenta $k$ since the loop momentum $p$ here becomes small. But as can be
shown including external momenta in the analysis cannot affect the
conclusion.

As a result for any diagram in perturbation theory one have the contribution
of the doubling states vanishing in the continuum limit.

From the other hand it is known that $SU(2)$ chiral YM model suffers from a
global anomaly \cite{witten}. As it was shown by Witten one cannot globally
define the sign of the fermionic determinant once topologically nontrivial
gauge transformations which can change the sign of the determinant are
considered.

In our case, however, it seems everything to be all right. One has gauge
invariant regularized action with doubler's contribution suppressed in the
continuum limit. And no gauge noninvariant effect like change of the
determinant sign can be produced here. The problem is that the above
constructed lattice regularization is equivalent to the model considered by
Witten only perturbatively.

To illustrate this consider eigenvalue problem for (modified) chiral Dirac
operator $\widehat{\mathcal{D}}=\left( \widehat{D}+\frac 1{2\Lambda ^2}\{
\widehat{D},\Delta \}\right) $. Charge conjugation relates both chiralities
in such a way that one can introduce scalar product such that chiral Dirac
operator is symmetric. Let $\{\lambda _i\}$ be the set of its eigenvalues.
Then one has for the following expression for the determinant (see ref.\cite
{sl3})
\begin{equation}
\det \widehat{\mathcal{D}}=\prod_i\frac{\left| \lambda _i\right| }{\lambda _i
}\tanh \left( \frac{\pi \left| \lambda _i\right| }{2M}\right)  \label{det}
\end{equation}
one can see from this that the large eigenvalues ($|\lambda _i|\gg M$) are
cut off but not their signs. The same happens with the contribution of the
doubling states. Arguments from the perturbation theory shall tell us that
$\frac{\left| \lambda _i\right| }{2M}\rightarrow \infty $ where $\lambda _i$
are eigenvalues corresponding to unwanted (doubling) states. But when they
are cut off they leave their signs alive in eq. (\ref{det}). From the other
hand from general reasons (e.g. Nielson--Ninomiya theorem) we know that
total number of the minus signs in eq. (\ref{det}) is an even one\footnote{
Moreover, due to hypercubic invariance it is $\propto 2^4=16$.}. This would
prevent the determinant from being negative.

Indeed, for a sufficiently small lattice spacing one can write down the
regularized determinant as follows (see Appendix)
\begin{eqnarray}
\det \widehat{\mathcal{D}} &\approx &\prod_i\frac{\left| \mu _i\right| }{\mu
_i}\tanh \left( \frac{\pi \left| \mu _i\right| }{2M}\right) \times
\label{a16} \\
&&\ \ \ \times \prod_{\text{\{doublers\}}}\prod_i\frac{\left| \mu _i\right|
}{\mu _i}\tanh \left( \frac{\pi \left| \mu _i\right| }{2M_{*}}\right) ,
\nonumber
\end{eqnarray}
where $\mu _i$ are eigenvalues of the continuum (chiral) Dirac operator and
$M_{*}$ is some effective cutoff mass of the order $M\Lambda ^2a^2\sim
\sqrt{a}$. We wrote the regularized Dirac operator determinant as a product
of determinants of 16 Dirac operators. Each of them goes to $\pm 1$ as $a$
goes to $0$ but their product is always $+1$.

This lead to vanishing of the Witten anomaly as it is seen from the lattice.
In fact on a lattice there is no reason for Witten anomaly since the
topology of the lattice is different from one of the continuum.

Now let us note that the cancelation of the phase factor here is identic to
the mechanism proposed by Slavnov in ref.\cite{sl3} but in this case
cancelation arises ``naturally'' from the lattice. The role of fields
introduced in ref.\cite{sl3} to cancel Witten anomaly is played here by the
lattice doubling states.

\section*{Discussion}

In the present work we proposed a chiral gauge invariant lattice
regularization of $SU(2)$ YM model with correct (perturbative) continuum
limit. The method includes using a second scale regularization with infinite
number of Pauli--Villars fields \cite{fs1} and insertion of a high order
derivative term into fermionic action.

There is no doubling or other lattice artefact on the perturbative level.
But despite vanishing perturbatively, the doubler's contribution produces
the Witten anomaly cancelation mechanism proposed in \cite{sl3}.

Another lecture we learned from this is that a lattice model which is
perturbatively equivalent to some continuum model can, however, be different
from it at nonperturbative level. Moreover, since the only way of defining
nongaussian path integral, beyond the perturbation theory is the lattice
discretization one is tempted to define nonperturbative features of the
continuum model as an extrapolation of the lattice one.

Since for a given lattice spacing the momentum function $P_\mu \left(
p\right) $ is bounded from above one can limit oneself to a finite number of
PV fields since the fields with $Mr\gg \frac 1a$ will decouple. In this case
decay of the integrand near the edge of the Brillouin zone will be
polynomial rather than exponential. The same arguments and conclusions,
however, will remain valid for this choice. But for practical calculation it
seems to be more convenient to use the Grassmannian even path integral where
the gauge invariant Wilson term can be constructed for calculation of
inverse (absolute value of) fermionic determinant.

\textbf{Acknowledgments: }I would like to thank Prof.A.A.Slavnov for useful
discussions, Pilar Hern\'andez for pointing my attention to ref. 
\cite{pilar}, and the Referee for suggestions leading to improvement 
of the manuscript. 

This work was supported by the RBRF under grant No 
96-01-00551.

\section*{Appendix}

In this appendix we will derive the expression (\ref{a16}) for the
determinant. Since the gauge fields $U_\mu $ are external ones for
sufficiently small lattice spacings $a$ they can be chosen close to the
unity. Also let $T^A$, $(A=1,...,15)$ be generators of the symmetry which
relates fermionic doubling states (see e.g. \cite{mak})
\begin{equation}
\{T^A\}=\{I,T^\mu ,T^\mu T^\nu \ (\nu >\mu ),...,T^1T^2T^3T^4\},  \tag{A1}
\end{equation}
where
\[
T^\mu =i\gamma _\mu \gamma _5(-)^{\frac{x^\alpha }a},
\]
commutes with the naive Dirac operator
\[
\widehat{D}T^A=T^A\widehat{D}
\]
This symmetry is responsible for fermion doubling. Laplace operator $\frac
1{\Lambda ^2}\Delta $ is not invariant under this symmetry\footnote{
This is also the reason why Wilsonian term kills fermion doubling states.}
but it transforms as follows
\begin{equation}
T_\alpha ^{-1}\Delta T_\alpha =\Delta +\frac 1{\Lambda ^2a^2}\left( U_\alpha
(x)\psi _{x+a\widehat{\alpha }}+U_{-\alpha }(x)\psi _{x-a\widehat{\alpha }
}\right)  \tag{A2}
\end{equation}
the second term under our conditions is approximately $\frac 1{\Lambda
^2a^2}\psi _x$.

Now consider the modified Dirac operator $\widehat{\mathcal{D}}=\left(
\widehat{D}+\frac 1{2\Lambda ^2}\{\widehat{D},\Delta \}\right) $ and
consider eigenvalues of $\widehat{\mathcal{D}}$ lying in the subspace of
small eigenvalues of $\Delta ,$ $\lambda (\Delta )\ll 1$. The spectrum of
the modified Dirac operator $\widehat{\mathcal{D}}$ here is close to the
spectrum of the naive one (and for small lattice spacings also to the
spectrum of the continuum Dirac operator). If now to act on such an
eigenstate with eigenvalue $\mu _i$ by a transformation $T$ the resulting
state will have eigenvalue close to $\frac 1{\Lambda ^2a^2}\mu _i$. This way
one finds fifteen such states. For greater values of $\widehat{D}$ this
relation becomes less and less exact but greater values are cut off and they
do not contribute (except the sign) to the determinant (\ref{a16}). The extra
factor $\sim \frac 1{\Lambda ^2a^2}$ can be included in the cut off mass by
its rescaling
\begin{equation}
M\rightarrow M_{*}=M\Lambda ^2a^2.  \tag{A3}
\end{equation}

Finally let us note that the requirement that $U_\mu \rightarrow 1$ as $
a\rightarrow 0$ is important here. This is natural since gauge fields are
external. But if one wants to consider dynamical gauge fields here one has
to introduce a regularization with a cut off of the order $\sim 1/\sqrt{a}$
also for the gauge fields. This can be easily seen from the perturbation
theory if one considers e.g. self-energy Feynman diagram for the fermionic
field. Since PV mechanism cannot be applied here one has to have a cutoff
for the gauge fields to get rid of doubling contribution of the fermionic
propagator.

\end{document}